# A low-temperature scanning tunneling microscope capable of microscopy and spectroscopy in a Bitter magnet at up to 34 T

W. Tao[1,2], S. Singh[1,2], L. Rossi[1,2], J. W. Gerritsen[2], B. L. M. Hendriksen[2,3],
A. A. Khajetoorians[2], P. C. M. Christianen[1,2], J. C. Maan[1,2], U. Zeitler[1,2], B. Bryant[1,2]

[1] *High Field Magnet Laboratory (HFML—EMFL), Radboud University, 6525 ED Nijmegen, The Netherlands*
[2] *Institute for Molecules and Materials, Radboud University, 6525 AJ Nijmegen, The Netherlands*
[3] *Materials & Structural Analysis, Thermo Fisher Scientific, Achtseweg Noord 5, 5651 GG Eindhoven, The Netherlands*

We present the design and performance of a cryogenic scanning tunneling microscope (STM) which operates inside a water-cooled Bitter magnet, which can attain a magnetic field of up to 38 T. Due to the high vibration environment generated by the magnet cooling water, a uniquely designed STM and vibration damping system are required. The STM scan head is designed to be as compact and rigid as possible, to minimize the effect of vibrational noise as well as fit the size constraints of the Bitter magnet. The STM uses a differential screw mechanism for coarse tip – sample approach, and operates in helium exchange gas at cryogenic temperatures. The reliability and performance of the STM are demonstrated through topographic imaging and scanning tunneling spectroscopy (STS) on highly oriented pyrolytic graphite (HOPG) at T = 4.2 K and in magnetic fields up to 34 T.

## I. INTRODUCTION

A Scanning Tunneling Microscope (STM) allows sample surfaces to be imaged with sub-nanometer topographic resolution, and enables local density of states to be directly probed via Scanning Tunneling Spectroscopy (STS)[1–3]. STM/STS at cryogenic temperatures and in magnetic fields have become crucial tools in condensed matter physics, enabling the study of unique physical phenomena such as atomic nanomagnets[4,5], surface states in topological insulators[6], and Landau quantization in low-dimensional electron systems[7–11]. Up to now, low temperature STM/STS in magnetic fields has been limited to a magnetic field strength of 18 T[12,13], as the majority of designs have been based on superconductor magnets. For some experimental applications, this field strength is not sufficient. For example, the study of fractal spectra in graphene superlattices requires at least 24 T to reach the 'critical' commensurability field where there is one flux quantum per superlattice unit cell[14,15]; observation of the room temperature quantum hall effect requires very high fields[16]; and many phase transitions, for example in some multiferroic materials, occur at fields above 20 T[17]. Static fields of more than 30 T can be generated in dedicated high-field facilities by water-cooled, resistive Bitter magnets, or hybrid resistive-superconducting magnets. However, implementing an STM in a Bitter magnet is a major challenge, due to the high level of vibrational noise produced by the turbulent cooling water, in addition to the strong space constraints resulting from the small magnet bore.

To date, only one STM has been implemented in such a magnet laboratory, namely the High Magnetic Field Laboratory in Hefei[18]. While the authors describe an STM which is operational in magnetic fields up to 27 T, the instrument cannot operate at cryogenic temperatures, severely limiting its application for high resolution STS.

In this paper, we present a specially designed STM capable of operating at cryogenic temperatures in a resistive Bitter magnet at the High Field Magnet Laboratory in Nijmegen. The instrument has been optimized for STS at 4.2 K. We review the STM head design, the cryogenic system and the vibration damping setup. We demonstrate atomic resolution imaging at 4.2 K, and high resolution STS in magnetic fields up to 34 T, on highly oriented pyrolytic graphite (HOPG). We observe a series of peaks in d$I$/d$V$ in high magnetic fields, which we attribute to Landau quantization of decoupled graphene as well as graphite.

## II. INSTRUMENT DESIGN

### A. Bitter Magnet

A schematic of the overall setup for STM measurements in high magnetic field is shown in Fig. 1. The STM has been designed to operate in a resistive Bitter magnet at the High Field Magnet Laboratory (HFML) in Nijmegen. Different types of Bitter magnets are available: for example two 33 T magnets and the larger 38 T magnet[19,20]. The Bitter magnets require water-cooling, with a typical flow rate of 140 l/s at a pressure of 20 bar: the high-pressure cooling water is supplied by a cooling plant. Both types of magnets have room temperature bores with a diameter of 32 mm:

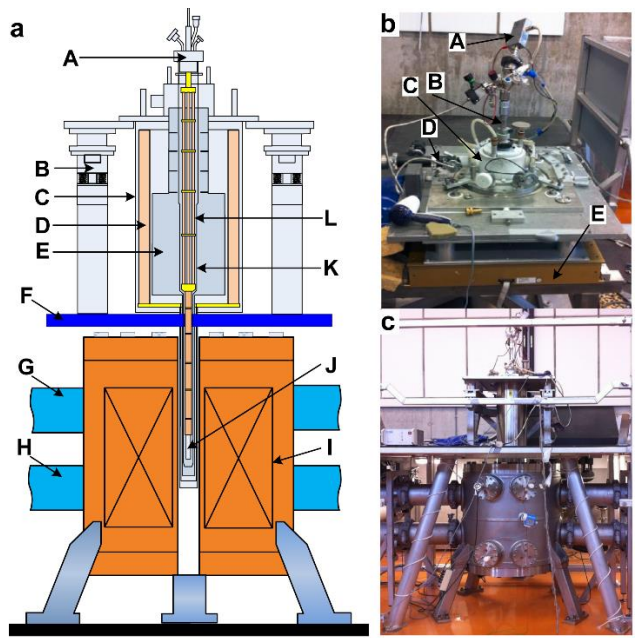

FIG. 1. (a) Schematic drawing of the whole high-field STM measurement setup. (A) STM probe, (B) active vibration dampers, (C) cryostat, (D) cryostat liquid nitrogen bath, (E) liquid helium bath, (F) support structure, independently mounted to the floor (G) incoming magnet cooling-water (H) outgoing cooling-water, (I) Bitter magnet coil (J) STM scan head, in magnetic field center (K) vacuum tube (L) STM mounting probe. (b) Image of topmost part of STM measurement system. The labeled parts are: (A) preamplifier circuit box, (B) STM insert, (C) cryostat, (D) bellow for helium recovery, (E) active vibration damper. (c) Image of the high field STM setup, in one of the 33 T Bitter magnets, showing the support structure.

this limits the sample space at cryogenic temperature to a maximum of 20 mm diameter. Helium cryostats with measurement probes are portable, and are separated from the magnet. The cryostat (Fig. 1(a), C) is mounted on two active vibration damping bars (B), on a support structure (F), which is mounted independently from the Bitter magnet (I) on the laboratory floor.

## B. Helium Cryostat

The STM probe (Fig. 1(a), A) is housed in a helium bath cryostat (C). The main body of the cryostat contains a helium bath (E) holding 10 L of liquid helium, shielded by a liquid nitrogen jacket (D). Both spaces are vacuum insulated. The hold time at helium temperature is typically 48 hours. The bottom part of the cryostat, the "tail", which is inserted into the bore of the Bitter magnet, is a long tubular cylinder with outer diameter 30 mm. The tail comprises a liquid helium space surrounded by a cylindrical copper radiation shield, which is inside the isolation vacuum. The copper shield is cooled by conduction from the liquid nitrogen in the main body of the cryostat. The inner diameter of the tail helium space is 24.2 mm: the experimental space at cryogenic temperature comprises a vacuum tube (K) inside the helium space, with an inner diameter of 20 mm. The vacuum tube is evacuated to a pressure of $10^{-5}$ mbar before cooling down: experiments are conducted in helium exchange gas.

## C. STM Scan Head and Insert

Two variants of the same design of STM head have been employed at HFML. The first is used in the 33 T Bitter magnets: this STM was originally used for experiments in superconducting magnets, its design is described in Dubois et al.[21]. The second is a new iteration of this design, used in the 38 T magnet. The major design constraints are that the STM head should be as rigid as possible, to minimize vibrational sensitivity, and it must be compact to fit inside the 20 mm diameter vacuum tube. It is also an advantage to be able to change tip or sample and cool down quickly, to maximize throughput in the limited experimental time available at the high field facility.

The 38 T STM scan head is illustrated in Fig. 2. Compared to the previous design[21] the STM head diameter has been reduced from 18 mm to 14 mm, affording 3 mm of clearance from the inner diameter of the vacuum tube. This makes it easier to avoid the STM touching the tube, which can cause vibrational noise and grounding issues. The STM head unit consists of three main parts. The first part comprises a differential screw used for coarse approach of the tip to the sample (Fig. 2(a), A). A sliding rod (C) is moved forward relative to the main barrel (D) by the differential screw (A). The differential screw mechanism is shown in an exploded view in Fig. 2(c).

The second component of the STM scan head is the scan unit consisting of a sliding plunger (Fig. 2(a), I). The piezoelectric scan tube (M) (outer diameter 6.35 mm, wall thickness 0.5 mm, and length 12.7 mm) is glued to the plunger using Epo-Tek H77 epoxy, with a Macor ring in between. The plunger is clamped by four leaf springs, which are made by taking cutouts of the mounting cylinder (Fig. 2(a), J). A hand-cut platinum/iridium tip (Pt80:Ir20) with diameter 0.25 mm is mounted in a titanium tip holder, which can be screwed into the end of the scan tube (O). The tip holder mounting is shielded from the scan tube voltages by a grounded shield. Coarse approach of the tip to the sample is effected by rotating the differential screw so that the pushing rod (F) slides the plunger forward on the leaf springs. The tip moves by 50 $\mu$m per complete turn of the differential screw: the total travel range is 1 mm. The spacer (G) allows for a mechanical decoupling between the plunger and the differential screw. After coarse approach, by retracting the differential screw two or three turns, the pushing rod can be decoupled from the plunger: this reduces the transmission of vibration to the tip-sample junction. A similar mechanism (a hook pair) was used in a room-temperature STM constructed for high magnetic fields, to decouple the STM from the piezo approach mechanism[18].

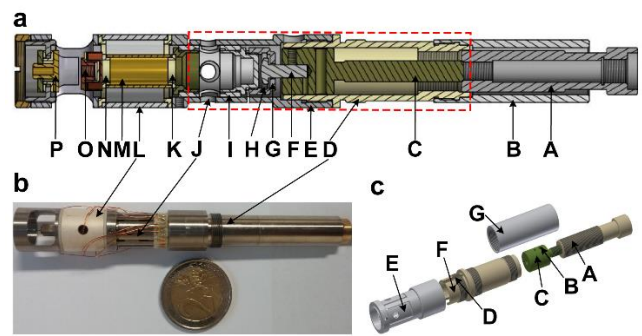

FIG. 2. 38 T STM scan head. (a) Cross-section drawing of STM head. The labeled parts are: (A) differential screw (B) protecting cylinder, (C) sliding rod, (D) main barrel, (E) cylinder with integral leaf springs, (F) pushing rod, (G) spacer to control the size of decoupling gap (H), (I) plunger on which the piezoelectric scan tube is mounted, (J) leaf springs, (K) and (N) insulating macor rings, (L) thermal compensation piezoelectric tube (M) piezoelectric scan tube, (O) tip holder mounting thread, (P) sample holder. (b) Image of STM scan head unit. (c) Exploded view of the coarse approach mechanism, as indicated by red dotted rectangle in (a). (A) Differential screw, (B) hole for screw to prevent rotation of the sliding rod, (C) sliding rod, (D) slot accommodating the fixing screw of the sliding rod, (E) leaf springs, (F) main barrel, (G) protecting cylinder.

The third part of the STM head accommodates the sample holder. A titanium end assembly is connected to the mounting cylinder (J) via a second piezoelectric tube with length 12.7 mm, width 14 mm, and thickness 0.8 mm (Fig. 2(a), L). This tube, which is of the same length and material as the scan tube, minimizes differential thermal contraction of the tip and sample assemblies. The end assembly has an inner thread, into which the sample holder mounting can be screwed. The sample holder mounting is fixed in place via a counter-nut: the inner thread in the end assembly is long enough to allow the sample holder mounting position to be varied. This enables tips of different length and samples of different thickness to be accommodated, whilst also minimizing the time taken for coarse approach. Finally the sample holder (P) is screwed into the mounting, from which it is insulated by two Macor inserts.

The STM scan head is fixed rigidly to a mounting probe comprising a tripod of glass fiber rods, (Fig. 1(a), L). The complete probe is inserted into a stainless steel vacuum tube (K). A long rod made of nonmagnetic stainless steel is employed to drive the differential screw. A stepper motor is used to drive the differential screw rod, enabling steps as small as 0.02° (translating to a tip movement of 2.5 nm) to be made.

Numerous measures have been taken to ensure the vibrational insensitivity of the STM. The compact design of the STM head provides a high resonant frequency: for the 33 T STM this is 9.2 kHz. The differential screw can deliver a great degree of force to the plunger, so the plunger can be strongly clamped by the leaf springs, contributing to the mechanical stiffness. All components of STM head are made of nonmagnetic titanium. The STM probe is constructed of nonmagnetic glass fiber rods, which provides some vibrational isolation due to its low resonant frequency and natural damping.

### D. Electronics

The Bitter magnet creates substantial electrical as well as mechanical noise. The power supply ripple is around 1 A, leading to a field ripple of 1 mT at 30 T. This could create a major problem for measuring tunnel currents of a few pA. However, the bias voltage and tunnel current lines are well shielded by the grounded cryostat and vacuum tube: the current preamplifier is mounted on top of the cryostat, where electromagnetic interference from the magnet is smaller, with a stray field less than 100 mT at full field.

To minimize microphonic pickup and further reduce electrical interference, miniature coaxial cables (solid copper conductor, Type C from Lakeshore Cryotronics) are employed for the bias voltage and tunneling current. Conventionally, the sample bias is applied to the sample (via a plug in the back of the sample holder) and tunnel current is measured on the tip. Braided pairs of twisted copper wires are used for piezoelectric scan tube signals, thermometry, and ground. All cables are wrapped tightly to the mounting tubes to minimize possible vibration-induced electromagnetic induction. The STM head can be separately grounded from the vacuum tube to check for mechanical contact and to aid in finding ground loops. In normal operation the STM, vacuum tube and cryostat are all commonly grounded.

The current preamplifier is a home-made design with a nominal gain of $10^9$ V/A. A commercial SPECS SPM control unit (Nanonis) is used to perform the control of the scan tube and data acquisition.

### E. Vibration isolation

For an STM working at in a Bitter magnet, the primary source of noise is vibrational noise produced by the highly turbulent flow of cooling water in the Bitter magnet. This noise is very broadband, essentially mimicking a white noise source. A secondary source of vibrational noise is the laboratory cooling water installation: the high-pressure pumps required to produce the cooling water flow produce noise peaks at frequencies below 50 Hz.

To decouple the STM as much as possible from the vibrational noise generated by the Bitter magnet and the cooling installation, we have a series of vibration isolation methods. Firstly, the support structure on which the STM setup is mounted forms a table over the Bitter magnet, and is separately fixed to the floor, without touching the magnet. The main support structure reaches to just above the magnet top plate (F in Fig. 1(a)): on top of this a secondary, upper support structure is mounted, which supports the cryostat via two active damping bars (Accurion Halcyonics VarioBasic). For the 38 T STM the upper support has been optimized to make it as rigid as possible. The upper support structure is required to be adjustable so that the cryostat can be accurately aligned in the magnet bore. The active damping is effective in the frequency range below 200 Hz: above this frequency damping is passive. Inside the cryostat there are further vibration isolation measures: the vacuum tube carrying the STM is kept centered in and decoupled from the cryostat helium bath by a ring of Be-Cu springs. Finally, the STM itself is kept from touching the vacuum tube, and the STM

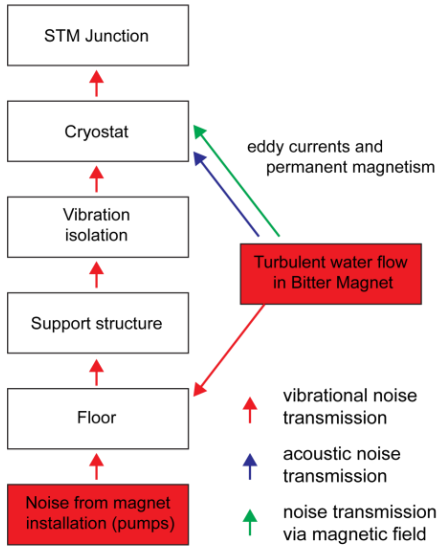

FIG. 3. Path of vibrational noise in the high-field STM system. Noise sources are shown in red: the path of vibration noise transmission is shown by red arrows, acoustic noise by blue arrows and noise coupling via the magnetic field by green arrows.

mounting tube forms a damped pendulum with a resonant frequency of a few Hz.

Some challenges remain with the vibration isolation system. One problem is that despite the relatively stiff support structure, which keeps the cryostat effectively mechanically isolated from the magnet, noise from the magnet can couple into the cryostat both acoustically and via the magnetic field, as the magnetic field is inhomogeneous compared to standard superconducting solenoids. The relative contribution of these two "short cuts" for vibrational noise will be seen in the following sections. Fig. 3 shows a schematic of the path of vibration transfer in the high field STM setup. The acoustic coupling is essentially unavoidable since there is only a 1 mm air gap between the magnet and the cryostat tail. Vibration of the magnet can couple into the cryostat via the magnetic field either due to permanent magnetism of the cryostat (which is largely made of stainless steel, and not completely nonmagnetic) or via eddy currents, since the vibrating magnet produces effectively a time-varying magnetic field. Vibration measurements indicate that eddy currents in the liquid nitrogen temperature copper shield in the tail of the cryostat contribute the bulk of the magnetic coupling between the Bitter magnet and the cryostat. A new cryostat designed to reduce the eddy current coupling is currently in progress.

### F. Operation of the high field STM

The procedure for setting up an experiment in the high field STM is as follows. The cryostat is mounted and aligned in the Bitter magnet. The sample and tip are loaded into the STM, and the STM into the vacuum tube. The tube is first flushed with helium gas, and then pumped with a turbo pump to $10^{-5}$ mbar. The vacuum tube is inserted into the cryostat: helium exchange gas is then introduced into the vacuum tube. The STM can be operated either at high pressures 100 mbar to 1000 mbar, or at low pressure < 0.1 mbar. Intermediate pressures are avoided due to the risk of Paschen discharge. The STM is normally operated at 4.2 K, but if desired a temperature down to 1.4 K can be attained by pumping the helium bath of the cryostat. Cool down to 4.2 K can be achieved in around one hour: samples and tips can thus be exchanged relatively rapidly, helping to achieve a high throughput of experiments. Coarse approach is effected by turning on the z-feedback with a small current setpoint, normally 5 pA, so that the scan piezo tube is fully extended. The stepper motor is then used to drive the differential screw and bring the tip into range of the sample. Once the tip is in tunneling range, the stepper motor is used to retract the differential screw two or three turns, so decoupling the screw from the plunger. STM imaging and STS can then be performed. Typically the tip is kept in tunneling range whilst ramping the field.

## III. PERFORMANCE

### A. Topographic imaging at 4.2 K

As an evaluation sample, we present STM imaging and STS spectroscopic data on Highly Oriented Pyrolytic Graphite (HOPG), which is cleaved via scotch tape exfoliation before being loaded into the STM.

Fig. 4 shows the effect of the Bitter magnet water flow and the magnetic field on the topographic imaging performance of the 33 T STM. The STM is operated inside a 33 T magnet at reduced water flow of 110 l/s, at which flow the maximum field is around 26 T. An HOPG sample is imaged: each image takes 4 minutes. Fig. 4(a) shows the effect of starting the cooling water flow whilst imaging. As can be seen from the image, the cooling water flow significantly reduces the image contrast, but atomic resolution can still be observed at 110 l/s. Fig. 4(b) shows the subsequent effect of ramping the magnetic field from 0 to 12 T, at 110 l/s, whilst imaging. Here the apparent effect is much more severe and atomic resolution is lost above 6 T. However, as can be seen in the image cross sections in Fig. 4(c), the noise in the z signal at around 10 T is still small, around 5 pm RMS (this increases to around 10 pm RMS at 24 T). This is substantially smaller than the atomic corrugation of 50 pm seen at zero water flow, implying that there must be substantially more noise in the x-y directions than in z. In fact at least a 200 pm broadening in the x-y plane is required in addition to the 5 pm z noise to obscure the atomic corrugation. This can be understood since noise will be amplified in the x-y directions by flexure of the piezo scan tube and STM body. In addition, the cryostat is suspended from the top on its active dampers, meaning that a pendulum motion can be set up: the dampers will be more effective at damping vibration in the z-axis vibration compared to the x-y axes.

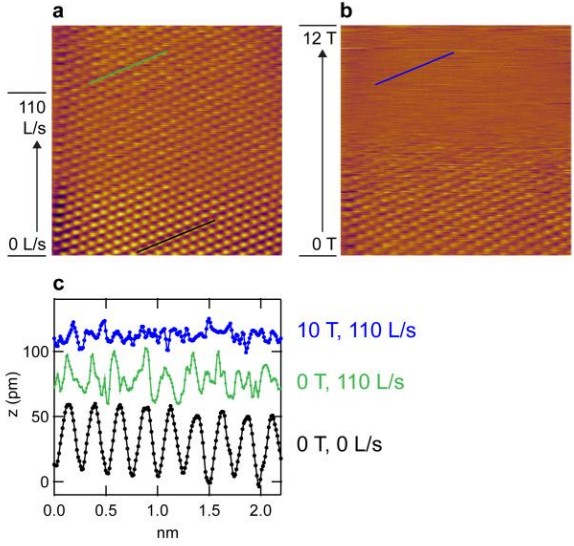

FIG. 4. Effect of Bitter magnet water flow and magnetic field on STM imaging with the 33 T STM (a) STM topographic image of HOPG, collected during the ramp up of the Bitter magnet water flow, 0 to 110 l/s (b) Imaging during field ramp from 0 to 12 T at 110 l/s. Both images 6 x 6 nm², 200 mV, 100 pA. (c) Sections through (a) and (b) showing noise and contrast under different conditions. Although atomic resolution is lost with the field ramp, the noise in the $z$ direction remains at approx. 5 pm RMS.

Although atomic resolution cannot yet be achieved at high fields, room temperature imaging tests indicate the $x$-$y$ resolution at 30 T is around 1 nm. Further, the noise in $z$ is small enough to suggest that the STM can be used effectively for scanning tunneling spectroscopy.

### B. Tunnel current noise and vibrational noise

Fig. 5 shows tunnel current noise in the 38 T STM as a function of field, with vibrational noise recorded on the cryostat for comparison. The STM is operated at 4.2 K in the 38 T Bitter magnet with an HOPG sample; the tip is kept in range (feedback on) during the field sweep from 0 to 32 T, at a water flow of 145 l/s, with a sample bias of 91 mV and tunnel current set point 200 pA. Vibrational noise is recorded simultaneously, via an accelerometer (Endevco 7754-1000) placed on top of the cryostat. Fig. 5(a) shows an FFT of the tunnel current, presented as a 2D plot as a function of time during the field sweep. The noise remains roughly constant up to 20 T, above which field there is a sharp rise in noise frequencies below 250 Hz. A similar result was observed in high-field dilatometry measurements in the 38 T magnet[22]. Fig. 5(b) shows tunnel current FFT spectra at zero field and 32 T, extracted from 5(a). The most prominent additional noise at 32 T is a broad band between 100 Hz and 250 Hz, and a peak at 41 Hz: this latter comes from the cooling installation high-pressure pumps. Although there is clearly some additional noise with field – deriving from the magnetic coupling of the Bitter magnet to the cryostat and STM assembly (Fig. 3), the noise increase is modest, with an RMS current noise at zero field of 5 pA, increasing to 8 pA at 32 T. No 50 Hz peak is seen in the tunnel current, suggesting that electrical interference is not a problem. Fig. 5(c) shows the vibration data, displayed in units of acceleration. Vibration spectra, acquired simultaneously with the tunnel current data in Fig. 5(b), are displayed for 0 T and 32 T. Additionally, a vibration spectrum recorded on top of the Bitter magnet is shown: the RMS noise up to 500 Hz is 50 times smaller on the cryostat than on the magnet. By comparing the accelerometer data to the tunnel current one may readily observe that the STM is more sensitive to low-frequency vibrational noise, for example the prominent vibrational noise peak at 290 Hz is absent from the tunnel current, but the peak at 41 Hz deriving from the high-pressure pumps is seen in both data sets.

Although the overall increase in vibrational noise from zero field to 32 T is moderate, there is relatively more additional noise at low frequencies, below 150 Hz. This increase and the low-frequency sensitivity of the STM accounts for the increase in noise in the tunnel current with field.

### C. Tunneling Spectroscopy at up to 34 T

Fig. 6 shows a sequence of d$I$/d$V$ STS recorded on HOPG in the 38 T magnet. We collected spectra at fields from zero to 32 T, with a water flow of 145 l/s: additionally in a separate experiment a spectrum was collected at 34 T, with a water flow of 154 l/s. The d$I$/d$V$ data are collected via a lock-in technique, employing an SR830 lock-in amplifier (Stanford Research Systems). An AC modulation signal of 1.5 mV RMS amplitude, 730 Hz is added to the DC bias voltage applied to the sample. The bias voltage was 250 mV, the current setpoint was 1 nA. Each spectrum has 250 points and takes around 5 minutes to collect. Good quality spectra are obtained at all fields: no increase in noise level was seen from 0 to 32 T, presumably thanks to the relatively high lock-in frequency, which avoids the increase in low-frequency noise. A slight increase in noise was observed when increasing to 154 l/s to enable a field of 34 T, but spectral features are still clearly visible. The spectrum at 34 T is taken with a different tip to the 0-32 T data, so cannot be directly compared, nonetheless many of the same spectral features may be observed.

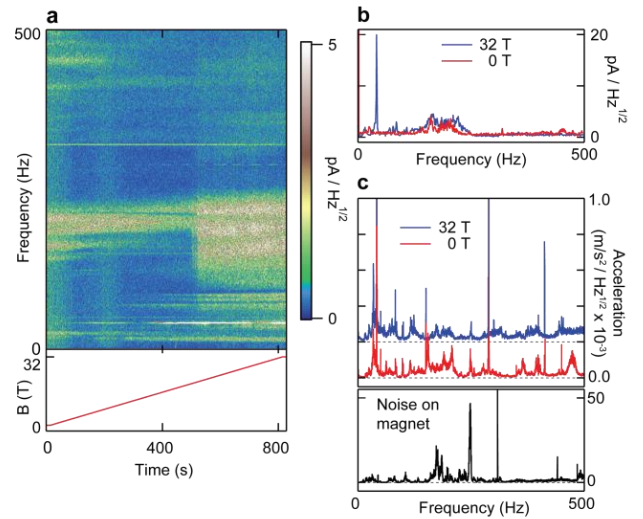

FIG. 5. (a) FFT of tunnel current noise in 38 T STM during a sweep from 0 T to 32 T, at 145 l/s cooling water. Sample bias of 91 mV, tunnel current setpoint 200 pA. (b) Individual tunnel current FFT spectra at zero field and 32 T. (c) Vibration spectra recorded on top of the STM cryostat. Spectra are shown for zero field and 32 T with 145 l/s water flow. Curves offset for clarity, zero lines are shown as dashed lines. The bottom panel shows a vibration spectrum recorded on the top plate of the Bitter magnet at 145 l/s water flow, the vertical scale is 100 x the other vibration plots.

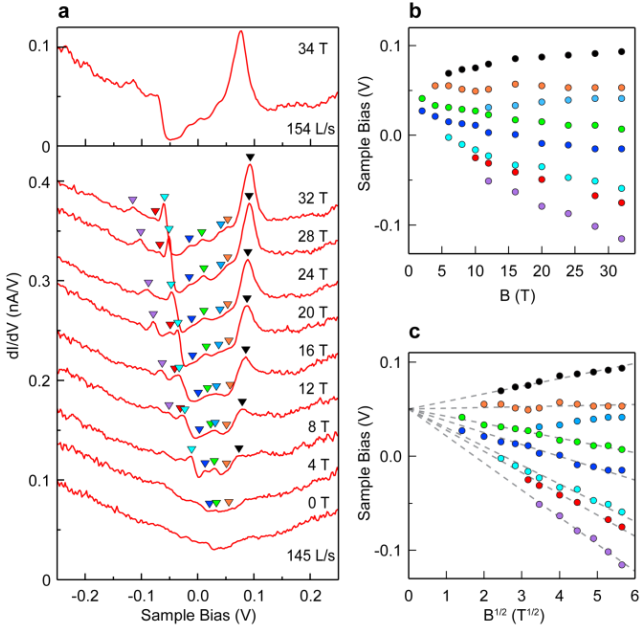

FIG. 6. (a) d$I$/d$V$ spectra of graphite, from the 38 T STM. One data set from 0 to 32 T, at 145 l/s water flow is presented, together with a single spectrum from a separate experiment at 34 T, 154 l/s. Spectra collected at 2, 6 and 10 T are omitted for clarity and curves are offset for clarity. (b) Peak positions from d$I$/d$V$ spectra up to 32 T plotted against field (B). (c) Same data plotted against B$^{1/2}$.

In Fig. 6 we observe peaks in the d$I$/d$V$ spectra at field, which evolve with field; these may be attributed to sample Landau levels. In order to accurately measure the peak energies (bias) we employ a second derivative analysis technique frequently used in infrared spectroscopy[23]. This allows us to remove the effect of the spectrum background, and identify nearby peaks which may otherwise not be clearly resolved. Fig. 6b shows the peak positions plotted against magnetic field, where each color indicates the evolution of a peak with field. The peak energies do not show a linear dependence on B, as would be expected for Landau levels of bulk graphite, but appear more linear with B$^{1/2}$ (Fig. 6 (c)); the Fermi energy is seen to be offset from zero bias by 50 mV.

A Landau level energy dependent on B$^{1/2}$ is characteristic of massless Dirac fermions in graphene[9,10]. However, our zero field spectra have nonzero d$I$/d$V$ at zero bias, contrary to the expectation for graphene of zero density of states at the Dirac point. Furthermore, we see more levels than would be expected for graphene in this bias and field range. It is likely that we observe simultaneously Landau levels both from bulk graphite and from surface graphene layers decoupled from the bulk, as has been observed previously for STS on graphite[9]. Weakly coupled graphene bilayers may also be present[24], which could explain the presence of extra Landau level peaks[25].

## IV. CONCLUSIONS

We have presented the design and characteristics of a scanning tunneling microscope, designed for cryogenic operation in a 38 T resistive Bitter magnet. We have demonstrated that it is possible to measure high-quality tunneling spectra at fields of up to 34 T, at 4.2 K, thanks to the highly rigid STM design, and vibration damping system.

This paper demonstrates the possibility of high resolution cryogenic ultra-high magnetic field scanning probe microscopy. Many challenges remain for the application of this new technique: operating an STM in user facility magnets is quite different from a conventional lab set up, since not only is the experiment time limited – and time at field is also limited by the availability of cooling water – but the whole setup must be portable, and set up anew for each experiment. Therefore a robust, reliable STM setup is paramount.

Several further developments are currently in hand. Firstly, so far atomic scale imaging at high magnetic fields has been prevented by the coupling of vibrational noise from the Bitter magnet via the magnetic field: eddy currents in the 77 K copper shield in the cryostat tail are primarily responsible for this coupling. A non-magnetic cryostat with measures to reduce the eddy current coupling is currently in progress, and it is hoped that this will allow atomic resolution at the full range of fields. Secondly, the existing STM design operates only in helium exchange gas, with no possibility of in-situ sample preparation. A full UHV or cryogenic vacuum dip-stick STM will create the possibility to study many different sample surfaces.

## ACKNOWLEDGEMENTS

The authors would like to thank Jos Rook (HFML) for design work and fabrication of the 38 T STM.

This work was supported by the Dutch funding organization FOM and by HFML-RU/FOM, a member of the European Magnetic Field Laboratory (EMFL).

A.A.K. acknowledges the VIDI project: 'Manipulating the interplay between superconductivity and chiral magnetism at the single atom level' with project number 680-47-534 which is financed by NWO.

**REFERENCES AND FOOTNOTES**

1. Binnig, G., Rohrer, H., Gerber, C. & Weibel, E. Tunneling through a controllable vacuum gap. *Appl. Phys. Lett.* **40,** 178–180 (1982).
2. Binnig, G., Rohrer, H., Gerber, C. & Weibel, E. 7 × 7 Reconstruction on Si(111) Resolved in Real Space. *Phys. Rev. Lett.* **50,** 120–123 (1983).
3. Chen, C. J. *Introduction to Scanning Tunneling Microscopy*. (Oxford University Press, 2007).
4. Hirjibehedin, C. F., Lutz, C. P. & Heinrich, A. J. Spin coupling in engineered atomic structures. *Science* **312,** 1021–4 (2006).
5. Khajetoorians, A. A. *et al.* Atom-by-atom engineering and magnetometry of tailored nanomagnets. *Nat. Phys.* **8,** 497–503 (2012).
6. Cheng, P. *et al.* Landau Quantization of Topological Surface States in Bi2Se3. *Phys. Rev. Lett.* **105,** 76801 (2010).
7. Niimi, Y., Kambara, H., Matsui, T., Yoshioka, D. & Fukuyama, H. Real-Space Imaging of Alternate Localization and Extension of Quasi-Two-Dimensional Electronic States at Graphite Surfaces in Magnetic Fields. *Phys. Rev. Lett.* **97,** 236804 (2006).
8. Luican, A., Li, G. & Andrei, E. Y. Quantized Landau level spectrum and its density dependence in graphene. *Phys. Rev. B* **83,** 41405 (2011).
9. Li, G. & Andrei, E. Y. Observation of Landau levels of Dirac fermions in graphite. *Nat. Phys.* **3,** 623–627 (2007).
10. Li, G., Luican, A. & Andrei, E. Y. Scanning Tunneling Spectroscopy of Graphene on Graphite. *Phys. Rev. Lett.* **102,** 176804 (2009).
11. Song, Y. J. *et al.* High-resolution tunnelling spectroscopy of a graphene quartet. *Nature* **467,** 185–189 (2010).
12. Song, Y. J. *et al.* Invited Review Article: A 10 mK scanning probe microscopy facility. *Rev. Sci. Instrum.* **81,** 121101 (2010).
13. Li, Q., Wang, Q., Hou, Y. & Lu, Q. 18/20 T high magnetic field scanning tunneling microscope with fully low voltage operability, high current resolution, and large scale searching ability. *Rev. Sci. Instrum.* **83,** (2012).
14. Dean, C. R. *et al.* Hofstadter's butterfly and the fractal quantum Hall effect in moiré superlattices. *Nature* **497,** 598–602 (2013).
15. Yu, G. L. *et al.* Hierarchy of Hofstadter states and replica quantum Hall ferromagnetism in graphene superlattices. *Nat. Phys.* **10,** 525–529 (2014).
16. Novoselov, K. S. *et al.* Room-Temperature Quantum Hall Effect in Graphene. *Science* **315,** 1379–1379 (2007).
17. Tokunaga, M. Studies on multiferroic materials in high magnetic fields. *Front. Phys.* **7,** 386–398 (2011).
18. Meng, W., Guo, Y., Hou, Y. & Lu, Q. Atomic resolution scanning tunneling microscope imaging up to 27 T in a water-cooled magnet. *Nano Res.* **8,** 3898–3904 (2015).
19. Wijnen, F. J. P. *et al.* Construction and performance of a 38 T resistive magnet at the Nijmegen High Field Magnet Laboratory. *IEEE Trans. Appl. Supercond.* **26,** 1–1 (2016).
20. Wiegers, S. A. J., Rook, J., den Ouden, A., Perenboom, J. A. A. J. & Maan, J. C. Design and Construction of a 38 T Resistive Magnet at the Nijmegen High Field Magnet Laboratory. *IEEE Trans. Appl. Supercond.* **22,** 4301504–4301504 (2012).
21. Dubois, J. G. A., Gerritsen, J. W., Hermsen, J. G. H. & Van Kempen, H. A low temperature scanning tunneling microscope for use in high magnetic fields. *Rev. Sci. Instrum.* **66,** (1995).
22. Küchler, R. *et al.* The world's smallest capacitive dilatometer, for high-resolution thermal expansion and magnetostriction in high magnetic fields. *Accepted to Rev. Sci. Instrum.* (2017).
23. Dong, A., Huang, P. & Caughey, W. S. Protein secondary structures in water from second-derivative amide I infrared spectra. *Biochemistry* **29,** 3303–3308 (1990).
24. Pereira, J. M., Peeters, F. M. & Vasilopoulos, P. Landau levels and oscillator strength in a biased bilayer of graphene. *Phys. Rev. B - Condens. Matter Mater. Phys.* **76,** 1–8 (2007).
25. Andrei, E. Y., Li, G. & Du, X. Electronic properties of graphene: a perspective from scanning tunneling microscopy and magnetotransport. *Reports Prog. Phys.* **75,** 56501 (2012).